\documentclass[aps,superscriptaddress,showpacs,nofootinbib,eqsecnum]{revtex4}

\usepackage{amsmath}
\usepackage{mathrsfs}
\usepackage{multirow}
\usepackage{hhline}

\usepackage{epsfig}  \input{epsf}
\def\be{\begin{equation}} \def\ee{\end{equation}} \def\bea{\begin{eqnarray}}
\def\eea{\end{eqnarray}}

\def\beq{\begin{equation}}
\def\eeq{\end{equation}}
\def\beqa{\begin{eqnarray}}
\def\eeqa{\end{eqnarray}}

\begin{document}

\title{The influence of a repulsive vector coupling in magnetized quark matter}

\author{Robson Z. Denke} \email{r.denke@posgrad.ufsc.br}
\affiliation{Departamento de F\'{\i}sica, Universidade Federal de Santa
  Catarina, 88040-900 Florian\'{o}polis, Santa Catarina, Brazil}

\author{Marcus Benghi Pinto} \email{marcus@fsc.ufsc.br}
\affiliation{Departamento de F\'{\i}sica, Universidade Federal de Santa
  Catarina, 88040-900 Florian\'{o}polis, Santa Catarina, Brazil}

\begin{abstract}
We consider  two flavor magnetized quark matter in the presence of a repulsive vector coupling ($G_V$) devoting special attention to the low temperature region of the phase diagram to show how this type of interaction counterbalances the effects produced by a strong magnetic field. The most important effects occur at intermediate and low temperatures affecting the location of the critical end point as well as the region of first order chiral transitions. When $G_V=0$ the presence of high magnetic fields ($eB \ge 10 m_\pi^2$) increases  the density coexistence region with respect to the case when $B$ and $G_V$  are absent  while a decrease of this region is observed at high $G_V$ values and vanishing magnetic fields. Another interesting aspect observed at  the low temperature region is that the usual decrease of the coexistence chemical value (Inverse Magnetic Catalysis)  at $G_V=0$ is highly affected by the presence of the vector interaction which acts in the opposite way. 
Our investigation also shows that the presence of a repulsive vector interaction enhances the de Haas-van Alphen oscillations which, for very low temperatures, take place at $eB \lesssim 6 m_\pi^2$. We observe that the presence of a magnetic field, together with a repulsive vector interaction, gives rise to a complex transition pattern since $B$ favors the appearance of multiple solutions to the gap equation whereas $G_V$ turns some metastable solutions into stable ones allowing for a cascade of transitions to occur.

\end{abstract}


\maketitle

\section{Introduction}

The  investigation of the effects produced by a  magnetic field ($B$) in the phase diagram of strongly interacting matter became  a subject of great interest in recent years. The motivation stems from the fact that strong magnetic fields may be produced  in non central heavy ion collisions \cite {kharzeev09}  (see  Ref. \cite {tuchin} for an updated discussion) as well as being present in magnetars \cite {magnetars}, and in the early universe \cite {universe}. At vanishing density both, model approximations \cite {eduardo, prd, newandersen, prd12} and lattice QCD evaluations \cite {earlylattice, lattice} agree that a cross over, which is predicted to occur in the the absence of magnetic fields \cite {Aoki},  persists when strong magnetic fields are present. However, a source of disagreement between recent lattice evaluations \cite {lattice} and model predictions regards the behavior of the pseudocritical temperature ($T_{\rm pc}$), at which the cross over takes place, as a function of the magnetic field intensity. The lattice simulations of Ref. \cite {lattice}, performed with $2+1$ quark flavors and physical pion mass values, predict that $T_{\rm pc}$ should decrease with $B$ while most model evaluations predict an increase (see Ref. \cite {prd} and references therein) which is also the outcome of  an early lattice evaluation performed with two flavors and high pion mass values \cite{earlylattice}.  
Possible explanations for the disagreement concerning how $T_{\rm pc}$ changes with $B$ have been recently given in Refs. \cite {prl,endrodi} but,  in any case, what seems sure is that one should expect that the high temperature-low baryonic density part of the QCD phase diagram be dominated by a cross over in the presence or in the absence of a magnetic field. The other extreme of the phase  diagram, which is related to cold and dense matter, is currently non accessible to lattice simulations so that one has to rely on model predictions. This region is especially important for astrophysical applications since the equation of state (EoS) for strongly interacting cold and dense matter is an essential ingredient in many  studies related to compact stellar objects. At vanishing magnetic field the EoS  obtained with most effective models \cite {scavenius, prcopt} have pointed out towards a  first order (chiral) phase transition taking place at a coexistence quark chemical potential  of the order of one third of the nucleon mass. As one moves towards intermediate values of $T$ and $\mu$ this first order transition line, which starts at $T=0$, will eventually terminate at a critical end point (CP) which separates the first order transition region from the cross over region. The precise location and nature of all the relevant phase boundaries  is model dependent and will also be influenced by the chosen parametrization as well as by the approximation adopted in the EoS evaluation. The question of how the first order transition line and the CP location would be affected by magnetic fields has been addressed in Ref. \cite {prd} in the framework of the three flavor NJL model. Regarding the high-$T$ and low-$\mu$ part of the phase diagram the results of Ref. \cite {prd} have only confirmed that the cross over location moves towards higher temperatures with increasing $B$ as predicted by other model applications \cite {eduardo} so that in this regime the broken symmetry region should expand as higher magnetic fields are considered in opposition to the recent lattice findings of Ref. \cite {lattice} as we have already remarked. On the other hand, one of the main results of Ref. \cite {prd}  shows that this  magnetic effect gets reversed in the low-$T$ and high-$\mu$ regime when the symmetry broken phase tends to shrink  with increasing values of $B$. Then, at low temperatures, the coexistence chemical potential value associated with the first order transition decreases with increasing magnetic fields in accordance with the Inverse Magnetic Catalysis phenomenon (IMC). This result has been previously observed with the two flavor NJL, in the chiral limit  \cite {inagaki}, as well as with  a holographic one-flavor model \cite {andreas} and more recently with the planar Gross-Neveu model \cite {novo}.  A model-independent physical explanation for IMC is given  in Ref. \cite {andreas} while a recent review with new analytical results for the NJL can be found in Ref. \cite {imc}. Another interesting result obtained in Ref. \cite {prd} concerns the size of the first order segment of the transition line which expands with increasing $B$ in such a way so that the CP becomes located at higher temperature and smaller chemical potential values. This result has been confirmed by a Functional Renormalization Group (FRG) application to  the two flavor meson-quark model  \cite {newandersen} suggesting that the presence of a magnetic field enhances the first order phase transition at intermediate temperatures. 
A further step towards understanding the influence of magnetic fields over the first order transition portion of the phase diagram was taken recently in Ref. \cite {prd12} which had, as one of its  main goals, the investigation of the phase coexistence region. With this aim the phase diagram for magnetized quark matter has been mapped into the $T-\rho$ plane (with $\rho$ representing the quark number density) where, for a given temperature, the mixed phase region is bounded by the high ($\rho^H$) and by the low ($\rho^L$) coexistent quark number densities. 
The results of Ref. \cite {prd12} show that for $eB \lesssim 9.5 \, m_\pi^2$ the high density branch of the coexistence phase diagram oscillates around its $B=0$ value as a consequence of filling the Landau levels which influences the values of quantities such as the latent heat. This finding may also have consequences regarding, e.g., the physics of phase conversion whose dynamics requires the knowledge of the EoS {\it inside} the coexistence region. For example, at a given temperature, the surface tension between the two coexisting bulk phases ($\rho^L$ and $\rho^H$) depends on the value of their difference \cite {jorgen,surten} an so is affected by the oscillations suffered by the coexistence boundary due to the presence of a magnetic field as recently demonstrated in Ref. \cite {andre}. 
Still regarding the low temperature portion of the phase diagram, which so far has been less explored than the high temperature portion, one notices that most applications consider effective models with scalar and pseudo scalar channels only, within the mean field approximation (MFA) framework. However, at  finite densities the effects of a chiral symmetric vector channel  may  become important \cite {volker,buballa, fuku08}. Regarding the QCD phase diagram it has been established that the net effect of a repulsive vector contribution, parametrized by the coupling $G_V$,  is to add a term $-G_V \rho^2$ to the pressure weakening the first order transition \cite {fuku08}. Indeed, it has been observed that the first order transition line shrinks, forcing the CP to appear at smaller temperatures,  while the first order transition occurs at higher coexistence chemical potential values as $G_V$ increases. It is important to note that this trend can be observed even if one does not consider explicitly a $G_V$ term at the classical (tree) level provided that the evaluations be carried beyond the MFA. Evaluations performed with the nonpertubative Optimized Perturbation Theory (OPT) at $G_V=0$ have shown that, already at the first non trivial order, the free energy receives contributions from two loop terms which are $1/N_c$ suppressed \cite {prcopt}.  It turns out that  these exchange type of terms, which do not contribute at the large-$N_c$ (or MFA) level, produce a net effect similar to the one observed with the MFA at $G_V \ne 0$. This is due to the fact that the OPT pressure displays a term of the form $-G_S/(N_f N_c) \rho^2$ where $G_S$ is the usual scalar coupling so that a vector like contribution can be generated by quantum corrections even when $G_V=0$ at the lagrangian (tree) level. The relation between the MFA at $G_V \ne 0$ and the OPT at $G_V=0$ and their consequences for the first order phase transition has been analyzed in great detail in Ref. \cite {ijmpe}. What is important to remark, for our present purposes, is that  a magnetic field and a repulsive vector channel produce opposite effects as far as the first order transition is concerned. Also, so far, the model investigations related to the influence of magnetic field on the high density part of the phase diagram have been carried out with the MFA neglecting the vector contribution despite its potential importance for dense matter. Here, our aim is to extend these previous applications by considering the contribution of a repulsive vector channel to the two flavor magnetized  NJL model within the MFA which, to the best of our knowledge, has not been considered before.  The work is organized as follows. In the next section we use the MFA to obtain the  free energy for the NJL with a vector interaction at finite $B,T$ and $\mu$. In Sec III we compare the phase diagrams on the $T-\mu$ and $T-\rho$ planes for the different scenarios which are produced by scanning over the values of $B$ and $G_V$. Our conclusions are presented in Sec. IV.

\section{The NJL Magnetized Free Energy with a vector interaction  }
Starting from the standard two flavor NJL model \cite{njl} one can consider the influence a vector interaction by considering the  the following Lagrangian density \cite {volker,buballa}

\begin{equation}
\mathscr{L} = \bar{\psi}(i \displaystyle \gamma_{\mu}\partial^{\mu}  -m)\psi + G_S[(\bar{\psi}\psi)^2+(\bar{\psi}i\gamma_5 \vec{\tau} \psi)^2]-G_V(\bar{\psi}\gamma^{\mu}\psi)^2,
\end{equation}
\noindent
where $G_V>0$ represents the corresponding repulsive vector coupling constant.
In the MFA the thermodynamical potential at finite $T$ and $\mu$ can be written in terms of the scalar condensate, 
$\langle {\bar \psi}\psi \rangle$, and the quark number  density, $\langle \psi^+ \psi \rangle$.
 Then, considering only the zeroth component of $G_V(\bar{\psi}\gamma^{\mu}\psi)^2$  one linearizes the interaction terms in the NJL density as 
 \begin{equation}
(\bar{\psi} \psi)^2 \simeq 2 \langle {\bar \psi}\psi \rangle \bar {\psi} \psi - \langle {\bar \psi}\psi \rangle^2 \,\,\,\,{\rm and}\,\,\,\, (\bar{\psi} \gamma^0\psi)^2 \simeq 2 \langle \psi^+ \psi \rangle \psi^+ \psi - \langle \psi^+ \psi \rangle^2 \,\,,
\end{equation}
where quadratic terms in the fluctuations have been neglected. Using this mean field approximation the Lagrangian density can be written as 
\begin{equation}
\mathscr{L} = \bar{\psi}(i \displaystyle \gamma_{\mu}\partial^{\mu}-M +\tilde{\mu}\gamma^0)\psi- \displaystyle{\frac{(M-m)^2}{4G_S}}+{\frac{(\mu-\tilde{\mu})^2}{4G_V}} \,\,\,,
\end{equation}
where the effective mass, $M$, and effective quark chemical potential, $\tilde \mu$, are determined upon applying
the corresponding minimization conditions $\delta \Omega/\delta M = 0$ and  $\delta \Omega/\delta \tilde{\mu} = 0$ to the thermodynamical potential.
Finally, performing the path integral over the fermionic fields the MFA  thermodynamical potential one gets  (see Ref. \cite {prcopt} for results beyond MFA)
\begin{equation}
\Omega=\frac{(M-m)^2}{4G_S}-\frac{(\mu-\tilde{\mu})^2}{4G_V}   +\frac{i}{2}{\rm tr} \int \frac{d^4p}{(2\pi)^4} \ln  [-p^2+M^2] \,\,.
\label{free}
\vspace{0.4 cm}
\end{equation}
\noindent  In order to study the effect of a magnetic field in the chiral transition at finite temperature and chemical potential a dimensional reduction is induced  via  the following replacements \cite {eduana}  in Eq. (\ref{free}):

\begin{equation}
p_0\rightarrow i(\omega_{\nu}-i\mu)\,\, , \nonumber
\end{equation}

\begin{equation}
p^2 \rightarrow p_z^2+(2n+1-s)|q_f|B \,\,\,\,\, , \,\,\mbox{with} \,\,\, s=\pm 1 \,\,\, , \,\, n=0,1,2...\,\,, \nonumber
\end{equation}

\begin{equation}
\int_{-\infty}^{+\infty}  \frac{d^4p}{(2\pi)^4}\rightarrow i\frac{T |q_f| B}{2\pi}\sum_{\nu=-\infty}^{\infty}\sum_{n=0}^{\infty}\int_{-\infty}^{+\infty} \frac{dp_z}{2\pi} \,\,,\nonumber
\end{equation}

\noindent where $\omega_\nu=(2\nu+1)\pi T$, with $\nu=0,\pm1,\pm2...$ represents the Matsubara frequencies for fermions, $n$ represents the Landau levels and $|q_f|$ is the absolute value of the quark electric charge ($|q_u|= 2e/3$, $|q_d| = e/3$ with $e = 1/\sqrt{137}$ representing the electron charge\footnote {We work in Gaussian natural units where $1 \, {\rm MeV}^2 = 1.44 \times 10^{13}\, G$.}).  Here, we work in the situation of chemical equilibrium so that $\mu_u=\mu_d=\mu$. Then, following Ref. \cite {prcsu2} we can  write the  thermodynamical potential as

\begin{equation}
\Omega ={\frac{(M-m)^2}{4G_S}}-{\frac{(\mu-\tilde{\mu})^2}{4G_V}}+{\Omega}_{vac}+\Omega_{mag}+{\Omega}_{med}\,\,\,,
\end{equation}
where the vacuum contribution to the effective potential is

\begin{equation}
{\Omega}_{vac} = -2N_cN_f\displaystyle\int {\frac{d^3{\bf p}}{(2\pi)^3}} \sqrt{p^2+M^2}.
\label{vac}
\end{equation}

As usual the divergent integral appearing in Eq. (\ref {vac}) can be regularized by a non-covariant sharp cut-off, $\Lambda$, yielding

\begin{equation}
{\Omega}_{vac} = {\frac{N_cN_f}{8\pi^2}} M^4 \left\{ \ln{\left [
{\frac{(\Lambda+\epsilon_{\Lambda})}{M}}\right ]-\epsilon_{\Lambda}\Lambda[\Lambda^2+{\epsilon_{\Lambda}}^2
]}\right\} \,\,\,,
\end{equation}
where $\epsilon_{\Lambda}$ represents the energy $\sqrt{{\Lambda}^2+M^2}$ at the cutoff momentum value $\Lambda$. The magnetic part of the thermodynamical potential is given by 

\begin{equation}
{\Omega}_{mag}=-\displaystyle\sum_{f=u}^d{\frac{N_c(|q_f|B)^2}{2\pi^2}}
\biggr\{  \zeta'[-1,x_f]-{\frac{1}{2}}(x_f^2-x_f)\ln{x_f}+{\frac{x_f^2}{4}}\biggl \}.
\end{equation}
In the last expression, we have used the definition $x_f= M^2/(2 |q_f| B)$ and the derivative of the Riemann-Hurwitz zeta function $\zeta'(-1, x_f) = d\zeta(z, x_f)/dz|_{z=-1}$ (see the appendix of \cite{prcsu2} for detailed steps). 
Finally, the last term ${\Omega}_{med}$ is the in-medium contribution to the effective potential

\begin{equation}
{\Omega}_{med} = -\displaystyle{\frac{N_c}{2\pi}} \sum_{f=u}^d \sum_{k=0}^{\infty}   \alpha_k(|q_f|B)\int_{-\infty}^{\infty} {\frac{dp_z}{2\pi}}
 \left\{T\ln{[1+e^{-(E_{p}+\tilde{\mu})/T}]}
+T\ln{[1+e^{-(E_{p}-\tilde{\mu})/T}]}\right\}\,\,\,,
\label{Omega med}
\end{equation}
where $\alpha_k = 2 - \delta_{k0}$ and $E_{p} = \sqrt{p_z^2+2k|q_f|B+M^2}$ with $|q_u|=2/3$ and $|q_d|=1/3$.  A similar expression for the magnetized thermodynamical potential at $G_V=0$ was  also obtained in  Ref.  \cite {klimenko} where  Schwinger's proper time approach has been used.
Solving $\delta \Omega/\delta M = 0$ and  $\delta \Omega/\delta \tilde{\mu} = 0$ we get the following coupled self consistent equations
\begin{equation}
 M = m-2G_S(\phi_{vac} + \phi_{mag} +\phi_{med})\,\,, 
 \label{meff}
\end{equation}
and
\begin{equation}
{\tilde \mu}= \mu-2G_V \rho
\,\,,
\label{mueff}
\end{equation}
where $\phi_{vac}, \phi_{mag}$, and $\phi_{med}$ respectively represent the vacuum, the magnetic and the in medium contribution to the scalar condensate, $\langle {\bar \psi} \psi \rangle$, while $\rho$ represents the net quark number density, $\langle { \psi}^+ \psi \rangle$. As pointed out in Ref. \cite {buballa}, $\tilde \mu$ is a strictly rising function of $\mu$. Note also that, in principle, one should have two coupled gap equations for the two distinct flavors: $M_u = m_{u} - 2G_S(\langle {\bar u} u \rangle + \langle {\bar d} d \rangle)$ and  $M_d = m_{d} - 2G_S(\langle {\bar d} d \rangle + \langle {\bar u} u \rangle)$ where $\langle {\bar u} u \rangle$ and  $\langle {\bar d} d \rangle$ represent the quark condensates which differ, due to the different electric charges. However, in the two flavor case, the different condensates contribute to $M_u$ and $M_d$ in a symmetric way and since $m_u=m_d=m$ one has $M_u=M_d=M$. The quantities $\phi$ and $\rho$ appearing in Eqs. (\ref {meff}) and (\ref{mueff}) are given by

\begin{equation}
\phi_{vac}=-
{\frac{MN_cN_f}{2\pi^2}}\left \{ \Lambda \epsilon_{\Lambda}
-\displaystyle {\frac{M^2}{2}}\ln{\left [ {\frac{(\Lambda+\epsilon_{\Lambda})^2}{M^2}} \right ]}
  \right \}\;\;,
\end{equation}
\begin{equation}
\phi_{mag}=-\displaystyle {\frac{MN_c}{2\pi^2}}\sum_{f=u}^d (|q_f|B)^2\left [ \ln{\Gamma(x_f)}-{\frac{1}{2}}\ln{(2\pi)}+x_f-{\frac{1}{2}}(2x_f-1)\ln{(x_f)}      \right ]\;\;,
\end{equation}
\begin{equation}
\phi_{med}=\displaystyle {\frac{MN_c}{2\pi}}\sum_{f=u}^d \sum_{k=0}^{\infty}   \alpha_k(|q_f|B)\int_{-\infty}^{\infty} {\frac{dp_z}{2\pi}}{\frac{1}{E_{p}}}\left [ n_{p}(\tilde{\mu},T)+\bar n_{p}(\tilde{\mu},T) \right ] \;\;,
\end{equation}
and 
\begin{equation}
\rho = \displaystyle{\frac{N_c}{2\pi}} \sum_{f=u}^d \sum_{k=0}^{\infty}   \alpha_k(|q_f|B)\int_{-\infty}^{\infty} {\frac{dp_z}{2\pi}}
 \left[ n_{p}(\tilde{\mu},T)-\bar n_{p}(\tilde{\mu},T) \right] \,\,,
\end{equation}
where
\begin{equation}
\begin{array}{ccc}
n_p (\tilde{\mu},T)= \displaystyle{\frac{1}{(1 + e^{(E_p - \tilde{\mu})/T})}} &{\rm and} & \bar{n}_p (\tilde{\mu},T)= \displaystyle{\frac{1}{(1 + e^{(E_p+ \tilde{\mu})/T})}}.
\end{array} 
\end{equation}

\noindent represent, respectively, the Fermi occupation number for quarks and antiquarks.
In order to better understand the low temperature results it is convenient to take the $T\to 0$ limit in the above equations since in this situation the integral over $p_z$ can be easily performed producing analytical results which facilitate the analysis of the numerical results.
At $T=0$ the relevant in-medium terms appearing in $\Omega$, $M$ and ${\tilde \mu}$ can be written as
\begin{equation}
{\Omega}_{med} = -\displaystyle{\frac{N_c}{4\pi^2}} \sum_{f=u}^d \sum_{k=0}^{k_{f,max}}   \alpha_k|q_f|B \Biggl \{ \tilde{\mu}k_F(k,B)-s_f(k,B)^2\ln{\left[ {\displaystyle\frac{\tilde{\mu}+k_F(k,B)}{s_f(k,B)}} \right]}\Biggr \}\theta(k_F^2)\;\;,
\label{Omega tau}
\end{equation}
\begin{equation}
\phi_{med}= \displaystyle\sum_{f=u}^d \sum_{k=0}^{k_{f,max}}\alpha_k {\frac{MN_c(|q_f|B)}{2\pi^2}}\ln{\left [ {\frac{\tilde{\mu}+k_F(k,B)}{s_f(k,B)}}    \right ]}\theta(k_F^2)\,\,,
\end{equation}
and 
\begin{equation}
 \rho = \sum_{f=u}^d \sum_{k=0}^{k_{f,max}} \alpha_k \frac{|q_f| B N_c}{2\pi^2}k_F(k,B) \theta(k_F^2) \,\,,
\label{eq_rho_t0}
\end{equation}
\noindent
where $k_F(k,B)$ represents the Fermi momentum,  $k_F=\sqrt{{\tilde \mu}^2-s_f(k,B)^2}$ , and $s_f(k,B) = \sqrt{M^2+2|q_f|kB}$.  The maximum number of Landau levels, $k_{f,max}$, needed to accommodate all states is given by 
\begin{equation}
k_{f,max} =\displaystyle \left \lfloor {\frac{\tilde\mu^2-M^2}{2|q_f|B}} \right \rfloor .
\label{kmax}
\end{equation}

\section{Numerical Results}

Before choosing our parameter values let us point out that although a vector term is known to be important at high densities in theories such as the Walecka model for nuclear matter its consideration is more delicate within a non renormalizable model such as the NJL where usually the integrals are regulated by a momentum cut-off, $\Lambda$. Within this model $G_S$, $m$, and $\Lambda$ are generally  fixed to reproduce the pion mass 
($m_\pi \simeq 135 \, {\rm MeV}$), the pion decay constant ($f_\pi \simeq 93\, {\rm MeV}$) and the quark condensate ($\langle {\bar \psi} \psi \rangle^{1/3} \simeq 250 \, {\rm MeV}$) which yields $\Lambda \sim 560-670 \, {\rm MeV}$,  $G_S \Lambda^2 \sim 2-3.2$ , and $m \sim 5 - 7\, {\rm MeV}$ (see Ref. \cite {buballa} for a complete discussion). In this work, we choose the set $\Lambda = 590 \,  {\rm MeV}$, $G_S \Lambda^2 = 2.435$ and $m = 6.0\,  {\rm MeV}$.  
However,  fixing $G_V$ poses an additional problem since this quantity should be fixed using the $\rho$ meson mass which, in general, happens to be higher than the maximum energy scale set by $\Lambda$. Then, $G_V$ is usually considered to be a free parameter whose estimated value ranges between $0.25 \, G_S$ and $0.5 \, G_S$ \cite {gv1,gv2} so that here we will vary this coupling between zero and $0.5\, G_S$. 
\begin{figure}[htp]
\centering
\includegraphics[width=0.4\textwidth]{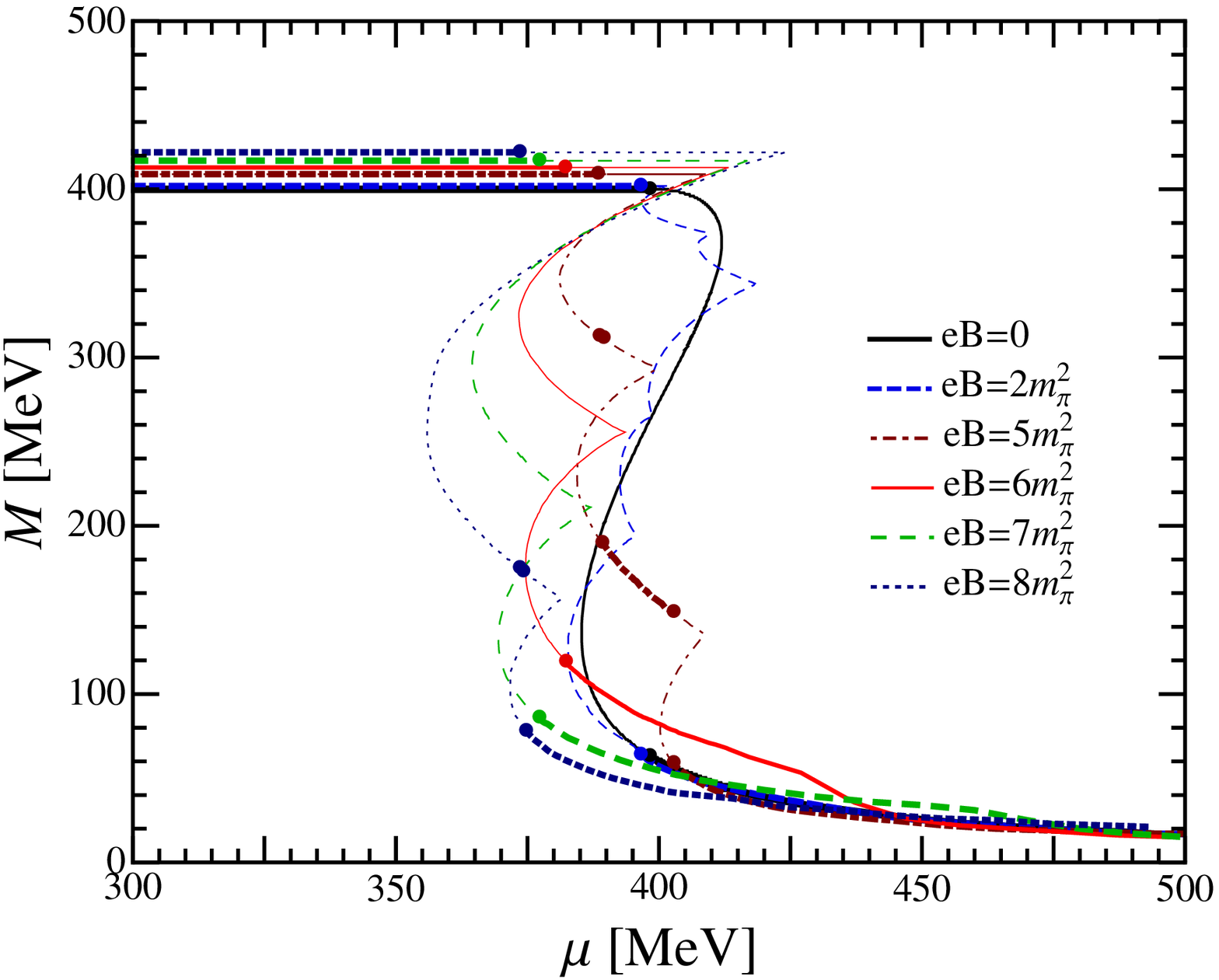} 
\includegraphics[width=0.4\textwidth]{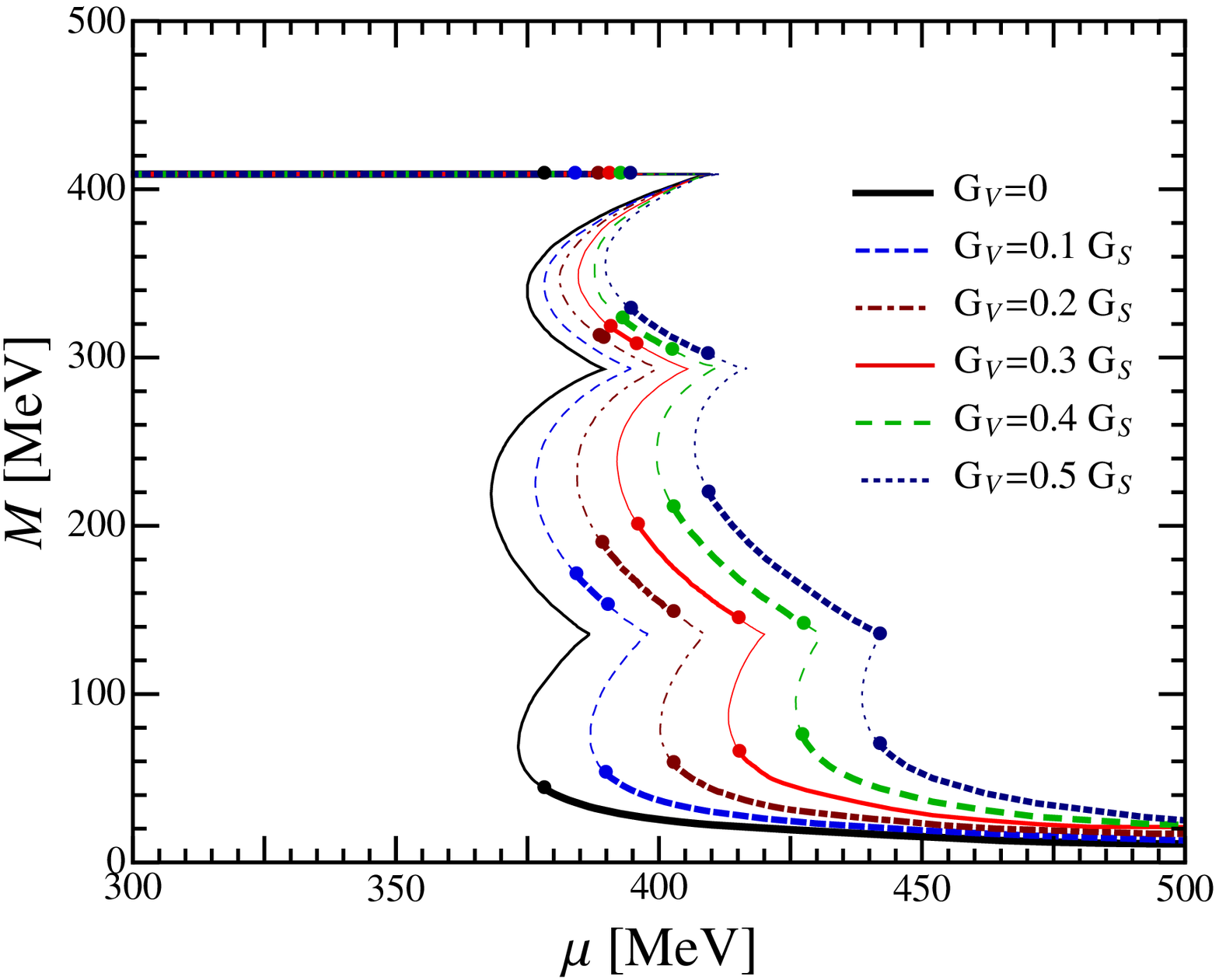} 
\caption{ Left panel: The effective quark mass, $M$, as a function of $\mu$ for different values of $eB$ at $G_V=0.2 G_S$. Intermediate transitions occur for $eB=5\, m_\pi^2$ and $eB=8\, m_\pi^2$. Right panel: The effective quark mass, $M$, as a function of $\mu$ for different values of $G_V$ at $eB=5\, m_\pi^2$. Intermediate transitions appear for $G_V \ne 0$ and become more pronounced as this coupling increases. In both cases the thick lines represent stable solutions to the gap equation.}
\label{fig1}
\end{figure} 

Usually, the order parameter associated with the chiral transition is taken to be $ (\langle {\bar u} u \rangle + \langle {\bar d} d \rangle)/2$ which, within the NJL model, can be related to the effective quark mass via $M=m- 2G_S (\langle {\bar u} u \rangle + \langle {\bar d} d \rangle)$. So, let us start by analyzing the effects of $B$ and $G_V$ over $M$ since this quantity also determines the behavior of the associated EoS. The left panel of Fig. \ref {fig1}  shows the effective quark mass, at $T=0$, as a function of $\mu$ for $G_V=0.2\, G_S$ and different values of the magnetic field.  The figure shows that, in the vacuum, the value of $M$ increases with $B$ which is in accordance with the magnetic catalysis effect \cite {MC}. Also, due to the filling of Landau levels, one observes the typical de Haas-van Alphen oscillations which are more pronounced for small values of $B$. Only the negative branches of $M(\mu)$ correspond to energetically favored gap equation solutions and in the present case ($G_V=0.2 \, G_S$) we observe that for $eB=5\, m_\pi^2$ and $eB=8\, m_\pi^2$ some of these solutions are stable leading to intermediate transitions. Each such a transition occurs at a given value of $\mu$ for which the gap equation has two stable solutions ($M^H$ and $M^L$) which then lead to the coexistence of two different densities (and therefore two different values of $\tilde \mu$) at the same pressure (and temperature).  Therefore,  a first order chiral transition will take place at some finite  (coexistence) chemical potential so that $P_{vac}(B,\mu=0,{\tilde \mu}=0,k_{f,max}=0,\rho^L=0,M^H)=P_{med} (B,\mu,{\tilde \mu},k_{f,max},\rho^H,M^L)$ where $\tilde \mu$, $M$ and $B$ are related through Eq. (\ref {kmax}). Then, as Fig. 1 shows, the value of $M^L$ at which   the vacuum pressure ($P_{vac}$) and the in medium pressure ($P_{med}$) are equal depends on the value of $B$. For example, at $G_V=0.2 \, G_S$ and $eB =5 \, m_\pi^2$ one sees that this  happens at $\mu=388.55 \, {\rm MeV}$ when $M^H=409\, {\rm MeV}$ and $M^L= 313 \, {\rm MeV}$ so that for these particular values of $B$ and $G_V$ (and also $\tilde \mu$ and $k_{f,max}$)  the  two degenerate minima of the thermodynamical potential are not too far apart. As one can see in Fig. 1 two subsequent first order phase transitions occur as $\mu$ increases until $M$ reaches the value $59\,{\rm MeV}$ at $\mu=402.65 \, {\rm MeV}$. Table I shows all the relevant values associated with this cascade of transitions. 
\begin{table}[htp]
\caption{\label{tab:table1}
Values assumed by $\mu$, $\tilde \mu$, $M$, $\rho_B$ (in units of $\rho_0= 0.17/{\rm fm}^3$), and $k_{f,max}$  at the three successive first order transitions occurring when   $eB=5 m^2_{\pi}$ and $G_V=0.2 G_S$.
}
\begin{ruledtabular}
\begin{tabular}{ccccccc}
$\mu$ [MeV] & $\tilde\mu$ [MeV] & $M$ [MeV] & $\rho_B/\rho_0$& $k_{u,max}$ & $k_{d,max}$ \\
\hline
\multirow{2}{*}{388.55} & 388.55 & 409.0 & 0     & 0 & 0  \\
{}                      & 379.6 & 313.0 & 0.82  & 0 & 0  \\
\hline
\multirow{2}{*}{389.05} & 379.9 & 312.0 & 0.83  & 0 & 0  \\
{}                      & 370.5 & 190.0 & 1.69  & 0 & 1  \\
\hline
\multirow{2}{*}{402.65} & 381.2 & 149.0 & 1.95  & 0 & 1  \\
{}                      & 373.8 &  59.0 & 2.63  & 1 & 2  \\

\end{tabular}
\end{ruledtabular}
\end{table}
If we now consider the case $eB=7\, m_\pi^2$, still at $G_V=0.2 \, G_S$, we see a different pattern in which only one transition occurs at $\mu= 377.5 \, {\rm MeV}$ when $M^H= 417 \, {\rm MeV}$ and $M^L= 85 \, {\rm MeV}$. 
For these values of $B$ and $G_V$ it is not energetically favorable for the thermodynamical  potential to have two degenerate global minima (stable solutions)  so close although a local minimum (metastable solution) occurs at $M=231 \, {\rm MeV}$ and $\mu= 377.5 \, {\rm MeV}$ as the figure shows. Table II shows all the relevant quantities for this phase transition and also for the metastable gap equation solution.

\begin{table}[htp]
\caption{\label{tab:table2}
Values assumed by $\mu$, $\tilde \mu$, $M$, $\rho_B$ (in units of $\rho_0= 0.17/{\rm fm}^3$), and $k_{f,max}$  at the first order transition  occurring when $eB=7 m^2_{\pi}$ and $G_V=0.2 G_S$. For reference we also show a metastable solution to the gap equation which occurs at $\mu=377.5 \, {\rm MeV}$ with the corresponding values of $\tilde \mu$, $M$, $\rho_B$, $k_{f,max}$ (all marked with a star). 
}
\begin{ruledtabular}
\begin{tabular}{ccccccc}
$\mu$ [MeV] & $\tilde\mu$ [MeV] & $M$ [MeV] & $\rho_B/\rho_0$ & $k_{u,max}$ & $k_{d,max}$ \\
\hline
\multirow{3}{*}{377.5} & 377.5 & 417.0 & 0     & 0 & 0 \\
{}                      & $361.2^*$ & $231.0^*$ & $1.48^*$  & $0^*$ & $0^*$   \\ 
{}                      & 351.4 &  85.0 & 2.38  & 0 & 1  \\

\end{tabular}
\end{ruledtabular}
\end{table}
The situation changes again for $eB=8\, m_\pi^2$ when two stable solutions occur for a high and an intermediate value of $M$ and as $\mu$ increases a little two degenerate minima occur at an intermediate and a low effective mass value. This trend of intermediate transition seems to be a result of the combined effects of $B$ and $G_V$ as should become clear by analyzing the right panel of the same figure where  we show the effective quark mass, also at $T=0$, as a function of $\mu$ for $eB= 5 \, m_\pi^2$ and different values of the vector coupling. At $G_V=0$ and $\mu =378.25\, {\rm MeV}$  the gap equation has seven solutions (two stable, two metastable, and  three unstable) and  only one transition occurs (between the two stable solutions, $M^H=409 \, {\rm MeV}$ and $M^L=43\, {\rm MeV}$)  but this simple pattern is highly affected by the presence of $G_V$ as the figure suggests.   As $G_V$ increases more intermediate transitions appear with the stable solution covering a higher range of chemical potential values. 
So, with increasing $G_V$ the thermodynamical potential develops degenerate global minima which are very close to each other and which will remain a stable solution over  a wider range of $\mu$ values. One clearly sees that as $G_V$ increases the first order chiral transitions become weaker and will eventually disappear for large values of this repulsive vector interaction.  The left panel of Fig. 1 suggests that the presence of a magnetic field favors the appearance of multiple solutions to the gap equation, especially at lower values of $B$ when the de Haas-van Alphen oscillations are more important. At the same time, the right panel of this figure shows that $G_V$ changes many of the metastable solutions into stable ones making the transition smoother.

\begin{figure}[htp]
\centering
\includegraphics[width=0.4\textwidth]{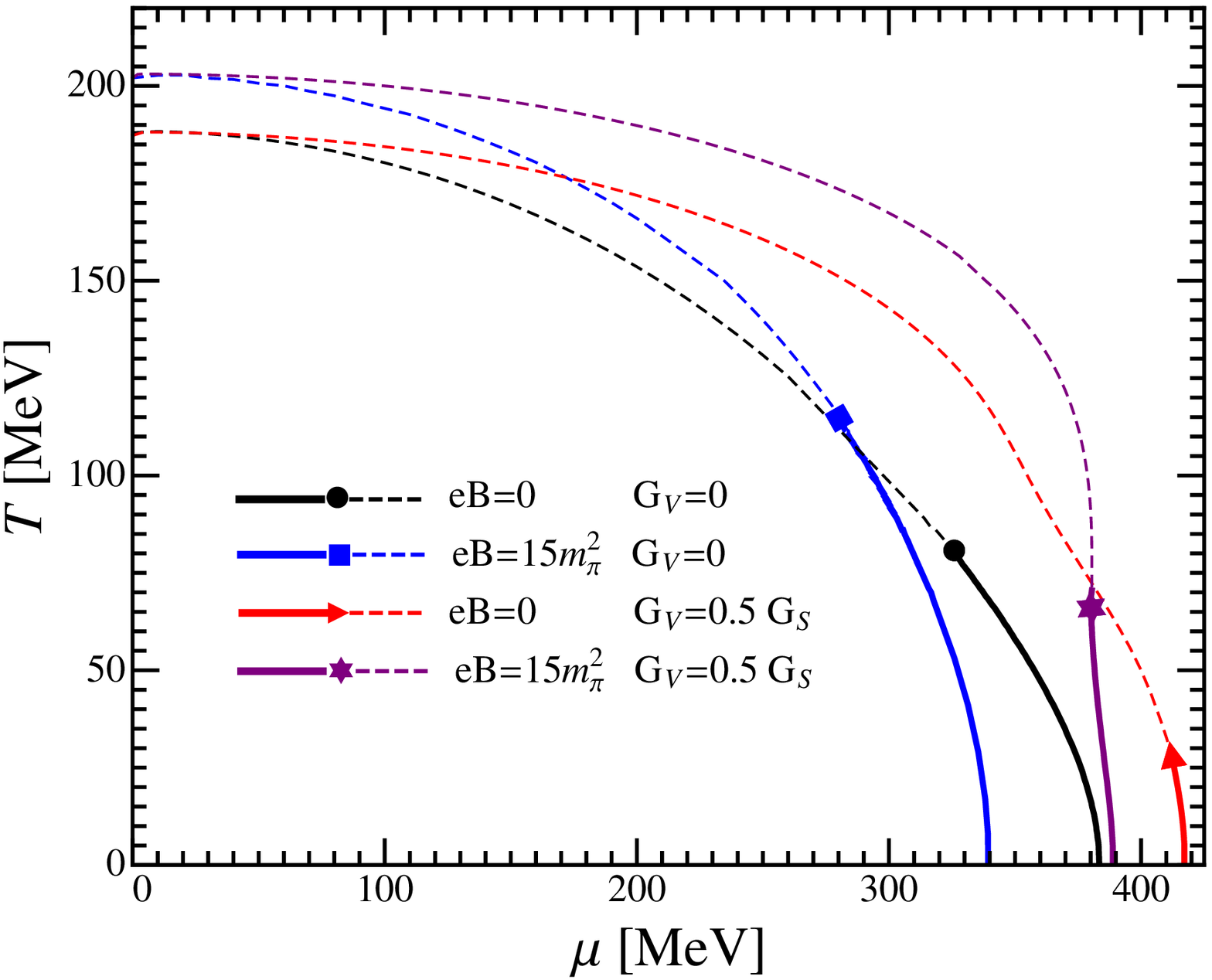} 
\includegraphics[width=0.4\textwidth]{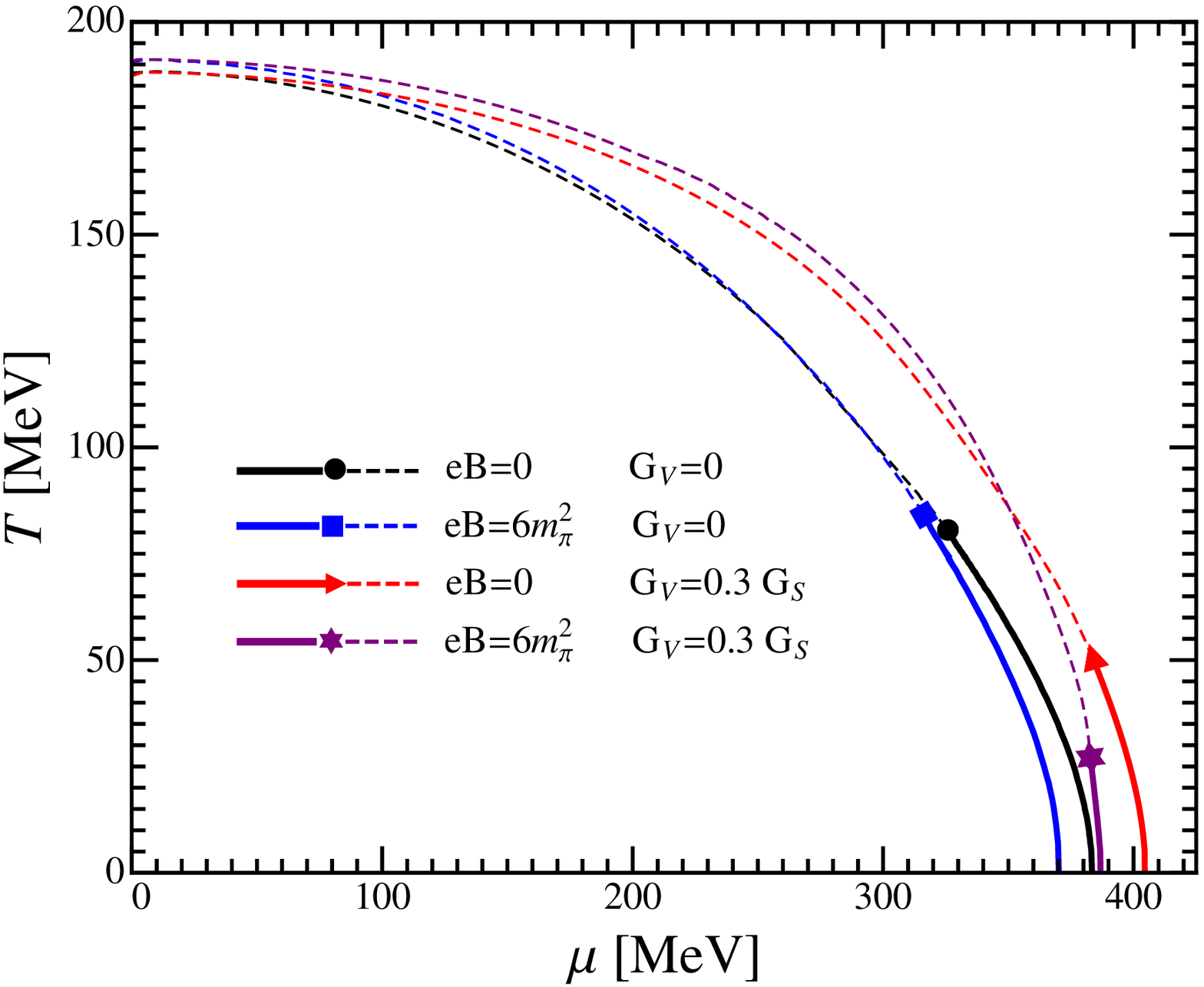} 
\label{PhaseDiagramB=15}
\caption{ Left Panel: Phase diagram in the $T-\mu$ plane showing the influence of the vector interaction ($G_V=0.5\,G_S$) and of the magnetic field ($eB=15$ $m_{\pi}^2$). Right Panel: Same as left panel figure but for $G_V=0.3\,G_S$ and $eB=6$ $m_{\pi}^2$. A continuous line represent a first order transition and s dashed line represents a cross overs while a solid symbol represents the CP.}
\label{fig2}
\end{figure}

\begin{figure}[h!]
\centering
\includegraphics[width=0.4\textwidth]{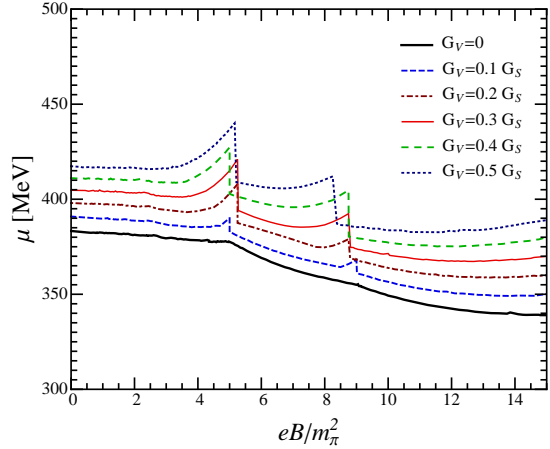} 
\caption{ Coexistence chemical potential as a function of $eB/m_\pi^2$,  at zero temperature, with varying vector coupling $G_V$. The vector coupling disfavors the decrease of the coexistence $\mu$ (IMC).}
\label{fig3}
\end{figure} 

\begin{figure}[htp]
\centering
\includegraphics[width=0.4\textwidth]{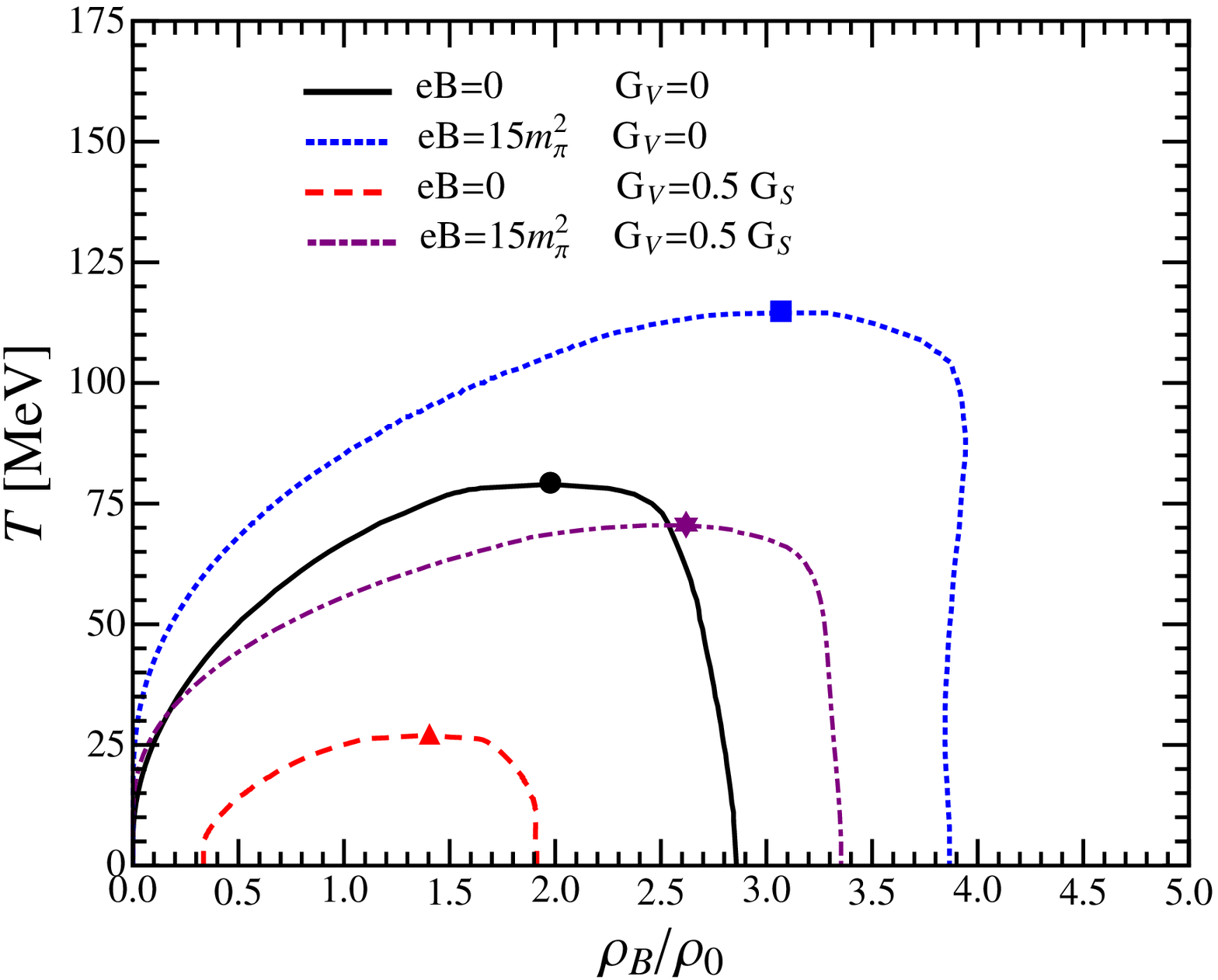} 
\includegraphics[width=0.4\textwidth]{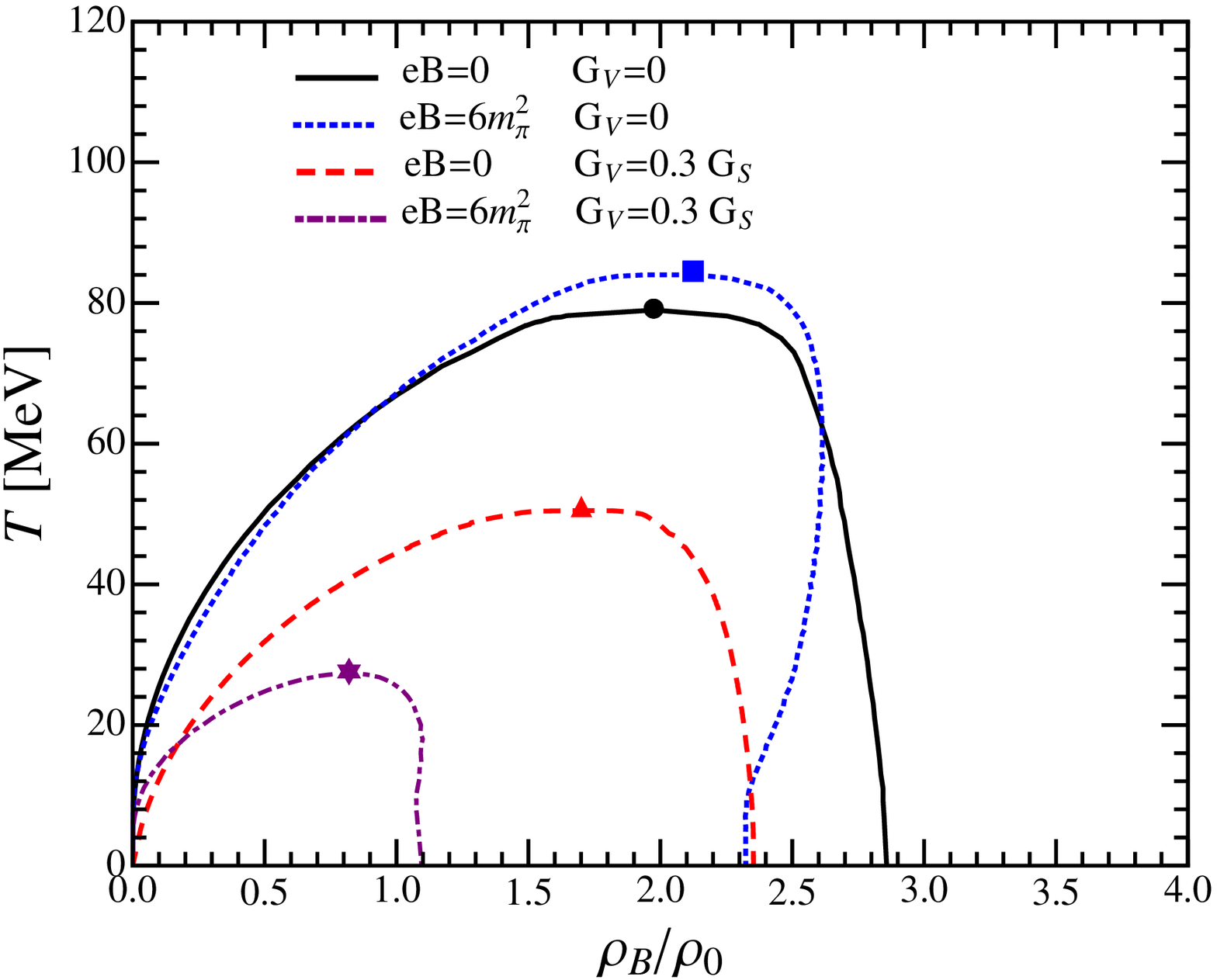} 
\caption{ Left Panel: Phase diagram for the first order region in the $T-\rho_B$ plane showing the influence of a vector interaction coupling of $G_V=0.5\,G_S$ and  of a magnetic field with intensity $eB=15\,m_{\pi}^2$. Right Panel: Same as the left panel figure but for $G_V=0.3 G_S$ and $eB=6\, m_\pi^2$.The baryonic density is given in units of $\rho_0=0.17/ {\rm fm}^{3}$.}
\label{fig4}
\end{figure}

\begin{figure}[htp]
\centering
\includegraphics[width=0.4\textwidth]{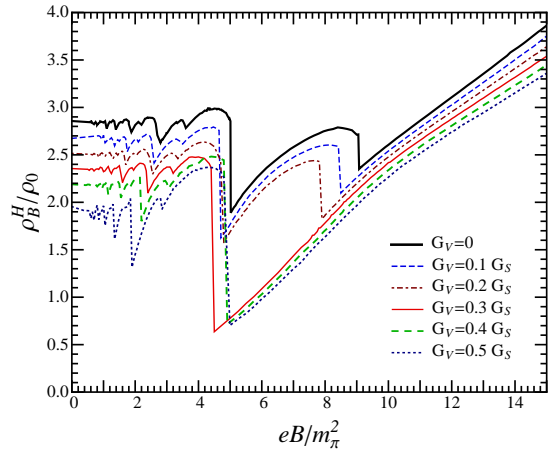}
\caption{ The high density branch of the coexistence region, $\rho_B^H$, as a function of $eB$ for $T=0$. Different values of $G_V$ have been considered to show the oscillation around the high density value at $B=0$. The baryonic density is given in units of $\rho_0=0.17/ {\rm fm}^{3}$. }
\label{fig5}
\end{figure}

\begin{figure}[htp]
\centering
\includegraphics[width=0.34\textwidth]{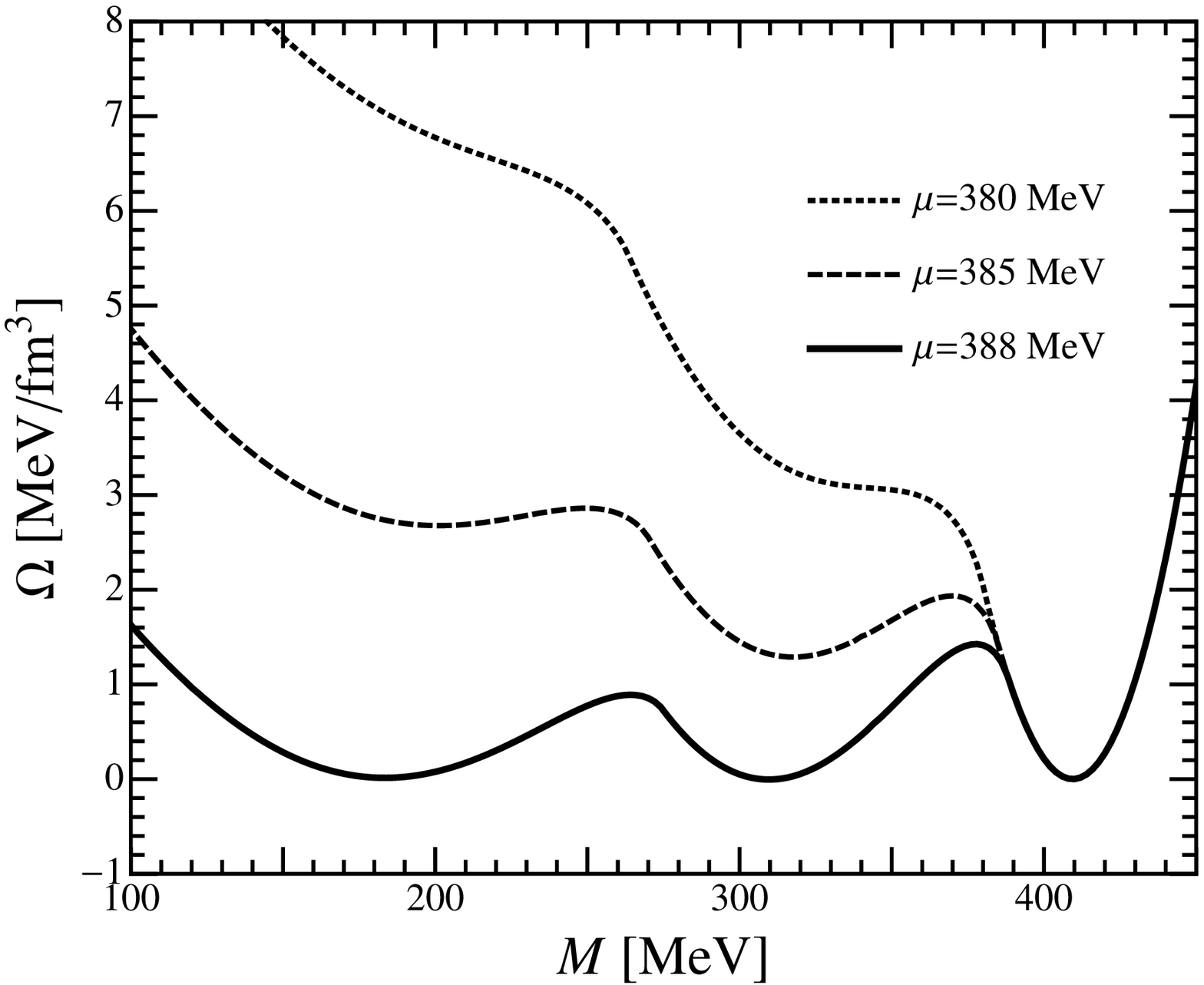} 
\includegraphics[width=0.35\textwidth]{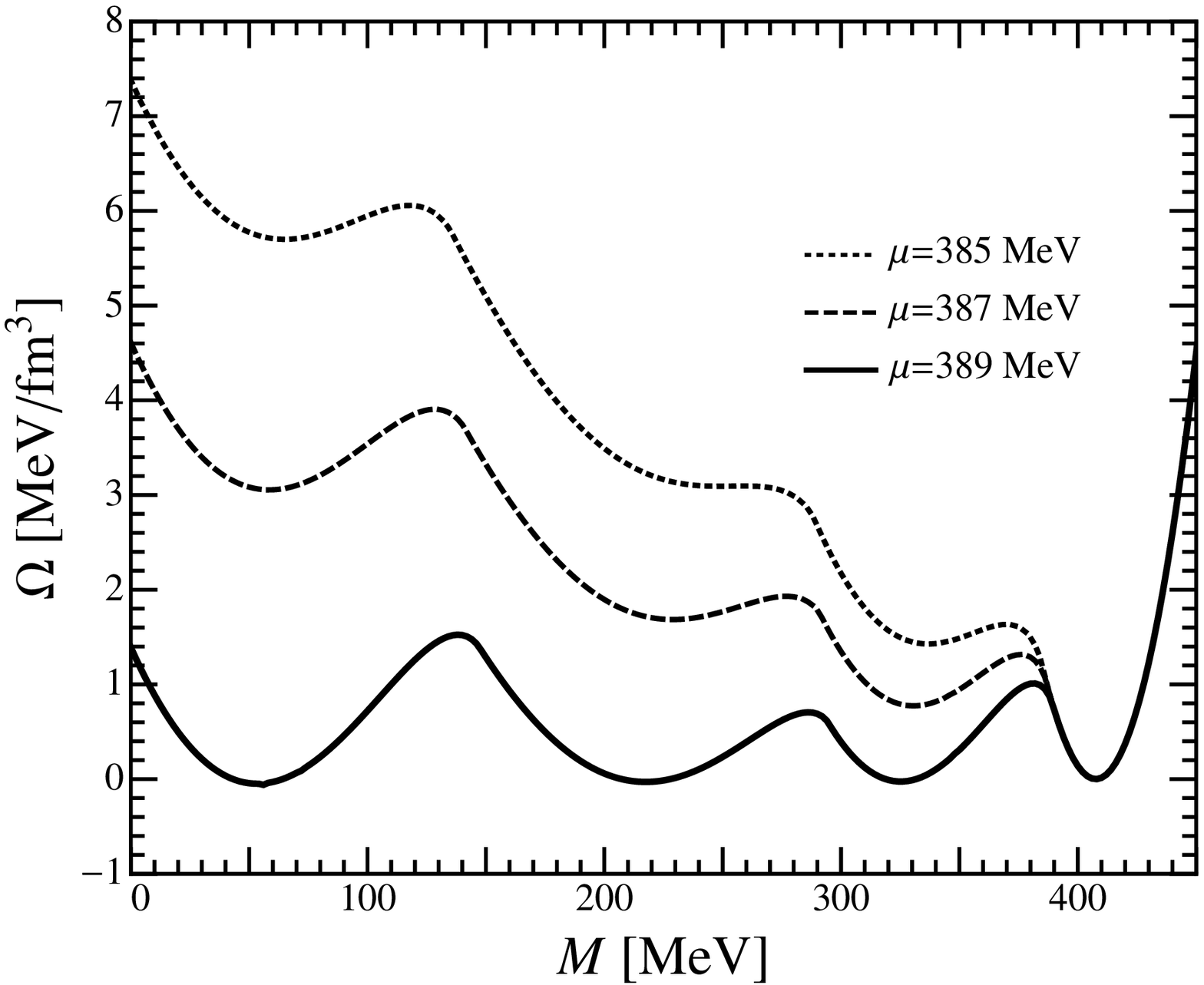} 
\caption{ Thermodynamical potential with multiple degenerate minima. Left panel: three degenerate minima occurring for $eB=5.1\, m_\pi^2$,  $G_V=0.2 G_S$ and   $\mu=388\,{\rm MeV}$. Right panel: four degenerate minima occurring for $eB=4.56\, m_\pi^2$,  $G_V=0.139 G_S$ and   $\mu=389\,{\rm MeV}$. }
\label{fig6}
\end{figure}

Let us now investigate how a repulsive vector interaction and a magnetic field affect the phase diagram of the two flavor NJL in the in the $T-\mu$ plane. For a given temperature, our criterion to select the coexistence chemical potential  when multiple first order transitions occur is to take the highest value of $\mu$ (after which $M$ decreases in a continuous fashion or will eventually go to zero if the chiral limit is taken).  For this numerical investigation let us consider the cases of $eB=6\, m_\pi^2$ and $eB=15\,m_\pi^2$ which are usually considered as representative values to be achieved at RHIC and the LHC respectively \cite {estimates}. Let us also set  $G_V=0.3\, G_S$, when analyzing  the $eB=6\, m_\pi^2$ case, and  $G_V=0.5 G_S$, when analyzing the $eB=15\,m_\pi^2$ case since these two particular coupling values are well suited to show how the repulsive vector interaction counterbalances the effects of the two magnetic fields values under investigation. The left panel of Fig. 2 shows the phase diagram for four situations described by $eB=0$ and $G_V=0$, $eB=15 \, m_\pi^2$ and $G_V=0$,
$eB=0$ and $G_V=0.5 G_S$ as well as $eB=15\, m_\pi^2$ and $G_V=0.5 G_S$. This figure clearly shows that the magnetic field enhances the first order chiral transition (case $eB=15 \, m_\pi^2$ and $G_V=0$)  while the repulsive vector interaction weakens this type of transition (case $eB=0$ and $G_V=0.5 G_S$).  The magnetic field alone also induces  a decrease of the value of the coexistence chemical potential associated with the first order phase transitions (Inverse Magnetic Catalysis \cite {andreas,imc}) while $G_V$ alone induces an increase of this quantity. However, their combined effect ($eB=15\, m_\pi^2$ and $G_V=0.5 G_S$) produces a less dramatic change with respect to the $eB=0$ and $G_V=0$ scenario, at least close to vanishing temperatures. The right panel shows that a similar situation occurs for  $eB=6 m_\pi^2$ and $G_V=0.3 G_S$. Comparing the two figures, at $T=0$,  we see that in both cases the coexistence chemical potential  for $eB \ne 0$ and $G_V \ne 0$ is also very close to the $eB = 0$ and $G_V = 0$ value.  But in the right panel,  for $eB=6\,m_\pi^2$ and $T\ne 0$, one observes that the combined effect of $B$ and $G_V$ is to weaken the first order transition even more than in the case of  $eB = 0$ and $G_V \ne 0$ so that the critical point appears at a very low temperature. As we shall see in the sequel, the reason for this difference can be traced back  to the de Haas-van Alphen oscillations. Before dealing with that  let us analyze  Fig. 3 which displays the coexistence chemical potential, at $T=0$, as a function of $B$ for different values of $G_V$. This figure clearly shows that the repulsive vector interaction inhibits the phenomenon of IMC apart from enhancing oscillations for $eB \le 8 m_\pi^2$.

To further understand the situation described in Fig. 2 let us now examine the $T-\rho_B$ plane ($\rho_B=\rho/3$) starting with the left panel of Fig. 4 for the case $eB=15\,m_\pi^2$ and $G_V=0.5 G_S$. Taking $B=0$ and $G_V=0$ as a reference, we see that $G_V$ alone tends to shrink the area of the coexistence region while $B$ alone works on the other way and  their combined effect is less severe as expected from the pattern observed in Fig. 2.  The right panel of Fig. 4 shows a similar type of figure for $eB=6\,m_\pi^2$ and $G_V=0.3 G_S$ showing that now the magnetic field alone (with $G_V=0$) does not produce a  dramatic increase of the coexistence region as in the previous case. Another interesting feature is that, at very small temperatures, the high density branch of the coexistence  region ($\rho_B^H$) for $eB = 6\, m_\pi^2$ and $G_V=0$ occurs at a lower value than for the $B=0$ case. So, comparing the $G_V=0$ curves of the left and the right panels of Fig. 4 one sees that $\rho_B^H$ oscillates around the $B=0$ value as originally observed in Ref. \cite {prd12}. Then, when $eB=6\, m_\pi^2$ and $G_V=0.3$ the decrease of $\rho_B^H$ is highly amplified so that the coexistence region area shrinks as a whole and as a consequence the critical end point appears at $T \simeq 30 \, {\rm MeV}$ as observed in the right panel of Fig. 2.  Figure 5 shows the high density branch of the coexistence region as a function of $B$ displaying  the oscillations which occur for $eB \lesssim 9.5\, m_\pi^2$. The oscillation pattern is such that one distinguishes pairs of similar cusps which occur, e.g., at $eB \approx 4 \, m_\pi^2$ and $eB \approx 9 \, m_\pi^2$ for $G_V=0$. This is due to the different charges of the {\it up} and {\it down} quarks so that, within this pair, the cusp at the lower $B$ value occurs when $k_{u,max}=1 \to 0$ for the {\it up} quark, which has the higher electric charge \cite {prd12}. For $eB \gtrsim 9.5 \, m_\pi^2$ only the lowest Landau level is filled so that there are no more oscillations and $\rho_B^H$ increases with $B$ in an almost linear fashion. The oscillations in $\rho_B^H$ can be further understood by analyzing the oscillations in  $M^L$ in conjunction with Eq. (\ref{eq_rho_t0})
and the interested reader is referred to Ref. \cite {prd12} for further details. 

So far we have observed (e.g. in Fig 2)  that a  cascade of transitions may occur  as a consequence of two degenerate minima appearing at successive values of $\mu$ for certain values of $B$ and $G_V$. However, even more exotic scenarios may emerge for other combinations of these two parameters. For example, by taking 
$eB=5.1\, m_\pi^2$,  $G_V=0.2 G_S$ and   $\mu=388\,{\rm MeV}$, at $T=0$, one observes three degenerate minima whereas by taking $eB=4.56\, m_\pi^2$,  $G_V=0.139 G_S$ and   $\mu=389\,{\rm MeV}$, at the same vanishing temperature, one observes four degenerate minima. These two situations are  illustrated in Fig. 6.

\section{Conclusions}
Using the MFA we have considered the magnetized two flavor NJL model with a repulsive vector interaction whose strength is given by the parameter $G_V$. Our aim was to concentrate in the less explored low temperature and high density region of the phase diagram for magnetized quark matter since this is the region where this type of interaction starts to play an important role. In the $T-\mu$ plane we  have observed that, 
in opposition to the magnetic field,  the repulsive interaction shifts the low temperature coexistence chemical potential to higher values counterbalancing the IMC effect observed at finite $B$ and vanishing $G_V$. In the $T-\rho_B$ plane we have seen that, due to the de Haas-van Alphen oscillations, the high density branch of the coexistence region associated to the first order chiral transition ($\rho_B^H$) oscillates  around the $B=0$ value for $eB \lesssim 9.5\, m_\pi^2$  and $G_V=0$. In the same situation, $\rho_B^H$ increases for $eB >9.5\, m_\pi^2$ producing results which are in accordance with the predictions of Ref. \cite {prd12}. The repulsive vector interaction on the other hand always favors the decrease of the value $\rho_B^H$ when $B=0$ so that  the combined effect of $B$ and $G_V$ will depend on their values. We have shown that for high fields (such as $eB=15\, m_\pi^2$) and high repulsion (such as $G_V=0.5 G_S$) the combined effect does not alter too much the $B=0$ and $G_V=0$ result while for a moderate field and coupling (such as $eB=6\, m_\pi^2$,$G_V=0.3 G_S$) we have observed that $\rho_B^H$, as well as  the whole coexistence region, suffers a substantial decrease whose origin  was associated to the oscillations suffered by $\rho_B^H$ due to the  filling of the Landau levels. We have  also concluded that the presence of a magnetic field, together with a repulsive vector interaction, gives rise to  complex transition patterns since $B$ favors the appearance of multiple solutions to the gap equation whereas $G_V$ turns some of the multiple metastable solutions into stable ones.  Then, depending on the values of  $B$ and  $G_V$ one may observe a cascade of first order transitions happening between two degenerate minima at different values of $\mu$. As $G_V$ increases the intermediate transitions are more frequent and cover a higher $\mu$ range weakening the first order transitions as expected from other studies (see, e.g. Ref. \cite {fuku08}). The interplay of $B$ and $G_V$ can produce even more exotic scenarios with multiple stable gap equation solutions arising at the same $\mu$ and $T$ for specific values of $B$ and $G_V$. In summary, our results show that the EoS for cold and magnetized quark matter can be highly affected by the inclusion a repulsive vector interaction which in turn may have consequences for the studies related to compact stellar objects such as magnetars. For example, very recently the surface tension for magnetized quark matter at $G_V=0$ has been evaluated \cite {andre} showing that the presence of an intermediate magnetic field (such as $eB \simeq 6\, m_\pi^2$) will reduce the value found at $B=0$ \cite {surten} which, in turn, would favor the presence of a mixed phase within magnetars \cite {veronica}. The results of the present work allows us to conclude that if a repulsive vector interaction is present the surface tension for magnetized quark matter value will be further lowered since (as seen in Fig. 4) this interaction contributes to the shrinkage of the coexistence region which is directly related to the surface tension value.

\section*{Acknowledgments}

 The authors  thank CNPq and Funda\c c\~{a}o de Amparo a Pesquisa e Inova\c c\~{a}o do Estado de Santa Catarina (FAPESC) for financial support.

\end{document}